\documentclass[%
preprint,
superscriptaddress,
amsmath,amssymb,
iop,longbibliography,
jcp,
floatfix,
]{revtex4-2}

\UseRawInputEncoding
\usepackage[utf8]{inputenc}
\usepackage{xcolor}
\usepackage{adjustbox}
\usepackage[colorlinks,allcolors=black,citecolor=blue,urlcolor=blue]{hyperref}
\usepackage{todonotes}
\usepackage{dirtytalk}
\usepackage{bm}
\usepackage{amsmath}
\usepackage{comment}
\usepackage[capitalise]{cleveref}
\usepackage{graphicx}
\usepackage{dcolumn}
\usepackage{dsfont}
\usepackage[T1]{fontenc}
\usepackage{mathptmx}
\usepackage{etoolbox}
\usepackage{xr}

\usepackage{amsmath}

\newcommand{\dint}{\textrm{d}}

\begin{document}

\title{Accurate nuclear quantum statistics on machine-learned classical effective potentials}

\author{Iryna Zaporozhets}
\affiliation{Department of Physics, Freie Universit\"at Berlin, Arnimallee 12, 14195 Berlin, Germany}
\affiliation{Department of Chemistry, Rice University, Houston, Texas 77005, United States}
\affiliation{Center for Theoretical Biological Physics, Rice University, Houston, Texas 77005, United States}

\author{F\'elix Musil}
\affiliation{Department of Physics, Freie Universit\"at Berlin, Arnimallee 12, 14195 Berlin, Germany}

\author{Venkat Kapil}
\email{v.kapil@ucl.ac.uk}
\affiliation{Yusuf Hamied Department of Chemistry,  University of Cambridge,  Lensfield Road,  Cambridge,  CB2 1EW, UK}
\affiliation{Department of Physics and Astronomy, University College London WC1E 6BT, UK}
\affiliation{Thomas Young Centre and London Centre for Nanotechnology, London WC1E 6BT, UK}

\author{Cecilia Clementi}
\email{cecilia.clementi@fu-berlin.de}
\affiliation{Department of Physics, Freie Universit\"at Berlin, Arnimallee 12, 14195 Berlin, Germany}
\affiliation{Center for Theoretical Biological Physics, Rice University, Houston, Texas 77005, United States}
\affiliation{Department of Chemistry, Rice University, Houston, Texas 77005, United States}

\begin{abstract}
The contribution of nuclear quantum effects (NQEs) to the properties of various hydrogen-bound systems, including biomolecules, is increasingly recognized. 
Despite the development of many acceleration techniques, the computational overhead of incorporating NQEs in complex systems is sizable, particularly at low temperatures. 
In this work, we leverage deep learning and multiscale coarse-graining techniques to mitigate the computational burden of path integral molecular dynamics (PIMD). 
Specifically, we employ a machine-learned potential to accurately represent corrections to classical potentials, thereby significantly reducing the computational cost of simulating NQEs. 
We validate our approach using four distinct systems: Morse potential, Zundel cation, single water molecule, and bulk water. 
Our framework allows us to accurately compute position-dependent static properties, as demonstrated by the excellent agreement obtained between the machine-learned potential and computationally intensive PIMD calculations, even in the presence of strong NQEs. 
This approach opens the way to the development of transferable machine-learned potentials capable of accurately reproducing NQEs in a wide range of molecular systems.
\end{abstract}
\maketitle %

\section{Introduction}
Molecular dynamics (MD) simulations provide insight into phenomena at atomic resolution and are an invaluable tool in chemistry, biophysics, and material science. 
Recent advances in force field~\cite{doi:10.1126/sciadv.adf0873,batatia2023foundation, VANDERSPOEL202118}, algorithms~\cite{https://doi.org/10.1002/jcc.26450} and hardware development~\cite{10.1145/3458817.3487397}  allow for increasingly higher accuracy and a broader scope of application of molecular dynamics simulations. \\

Traditional MD simulations rely on the classical evolution of nuclear positions, neglecting an explicit treatment of so-called quantum nuclear effects (NQEs) such as zero-point fluctuations, tunneling phenomena, and isotope effects. 
This simplification has a historical justification as the parameters of classical empirical force field simulations are fit to reproduce (quantum) experimental data. 
Hence, NQEs are effectively included in the classical potential energy surface, at least partially, albeit with limited transferability.
On the other hand, bottom-up \textit{ab initio} MD methods or machine learning interatomic potentials (MLIPs), based on the Born-Oppenheimer approximation~\cite{BornOppenheimer}, only account for the quantization of electrons. 
Hence, these simulations require an explicit treatment of quantum nuclear motion on the Born-Oppenheimer potential energy surface (PES) in theory.
Particularly when accurate electronic structure methods are used to estimate the Born-Oppenheimer PES, the neglect of quantum nuclear motion may be the largest source of error for obtaining quantitative agreement with experiments~\cite{medders_dissecting_2016, marsalek_quantum_2017, Pereyaslavets2018, daru_coupled_2022}. 
Unfortunately, NQEs are typically neglected due to their significant computational overhead or because they are assumed to have small effects.
However, even at ambient temperatures, NQEs significantly affect properties of hydrogen-bounded systems, e.g. static and dynamic properties of water~\cite{Stern2001, Shiga2005, Paesani2006, Morrone2008, Habershon2009, Vega2010, Pamuk2012, Fritsch2014, Wang2014, Litman2017, shepherd_efficient_2021, Eltareb2023, kapil_first-principles_2023}, free energies of hydrogen bonding and proton transfer~\cite{Tuckerman1997, Walker2010, Ceriotti2013, Fang2016, kapil_assessment_2019, kapil_complete_2022}, stability of molecular crystals~\cite{10.1063/1.5123992, doi:10.1021/jacs.1c10885}, strength of intermolecular interactions~\cite{Sauceda2021, Pereyaslavets2018}, and enzyme activity~\cite{doi:10.1073/pnas.0911416106, Wang2014}. 
\\

The standard condensed-phase simulation technique for incorporating NQEs in equilibrium properties is Feynman's imaginary time path integral (PI) formalism~\cite{feynman1965}. 
In this framework, a system's quantum statistical mechanics is mapped onto the classical statistical mechanics of a ring polymer, which consists of multiple replicas of the system connected by harmonic interactions that depend on temperature and mass. 
The effective classical PI partition function of the ring polymer system can be sampled using MD or Monte Carlo techniques, yielding the PIMD~\cite{10.1063/1.441588} and PI Monte Carlo~\cite{doi:10.1146/annurev.pc.37.100186.002153} methods.
In theory, PI methods incorporate quantum statistics exactly with classical sampling, in the limit of the number of replicas going to infinity~\cite{TuckermanBook}.
However, typically, the required number of replicas to converge structural properties, and thus the increased computational cost compared to classical MD, rises with increasing physical frequencies or lowering temperatures~\cite{markland_nuclear_2018}. 
As a result, most molecular systems require $16 - 32$ replicas at room temperature, which increases proportionally with lowering temperature, such as over $10^4$ replicas to obtain converged energies for molecular systems at cryogenic temperatures~\cite{Uhl2016}. \\

Over the last decades, several systematically improvable reduced-cost PI approaches have been proposed. 
These include high-order factorization of the Boltzmann operator~\cite{Takahashi1984, CHIN1997344, Perez2011, kapil_high_2016, kapil_modeling_2019, Kamibayashi2016}, , perturbative expansion of the Boltzmann operator~\cite{poltavsky_accurate_2020, poltavsky_modeling_2016} , multiple time stepping~\cite{kapil_accurate_2016}, ring polymer contraction~\cite{MARKLAND2008256}, and coloured-noise thermostats~\cite{PhysRevLett.103.030603, Dammak2009, ceriotti_efficient_2012}. 
Despite their notable success towards mainstreaming the use of PI methods~\cite{markland_nuclear_2018}, these approaches do not yet present a ``fix-all" solution. 
Issues still exist, including, but not limited to, lack of universal applicability of multiple time stepping and ring polymer contraction, tedious implementation of high-order path integral techniques, and poor ergodicity~\cite{korol_cayley_2019} or zero-point energy leakage~\cite{PhysRevLett.103.030603} due to the coupling between physical and fictitious PI collective modes.   
Finally, these approaches do not eliminate the diverging cost of path integral simulations at low temperatures. \\

An appealing direction that avoids the computational overhead of PI methods is to avoid modeling the entire imaginary-time path. 
For instance, the Wigner-Kirkwood effective potential~\cite{Wigner1932, Kirkwood1933} expands the partition functions in powers of the reduced Plank's constant and retains computationally tractable terms. 
Similarly, an effective quantum potential was suggested by Feynman and Hibbs \cite{feynman1965, feynman1972} by integrating out all degrees of freedom except the centroid of the ring polymer, i.e., by coarse-graining the ring polymer to the centroid.
Ever since, the idea of ring polymer coarse-graining has been explored for over 20 years~\cite{10.1063/1.1392355, 10.1063/1.1407291,  https://doi.org/10.1560/V0M8-VJPP-6Y31-BNFC} and has recently gained increasingly more attention thanks to the advent of MLIPs~\cite{Musil2022, doi:10.1021/acs.jctc.2c00706, doi:10.1021/acs.jctc.3c00921}. \\

A number of coarse-grained effective potentials can be developed by identifying (a) a suitable mapping from the fine-grained ring polymer to the coarse-grained classical system and (b) the functional form of the interactions in the coarse-grained system. 
For instance, the so-called ``CG-PI theory", developed by Voth \textit{et al.}, \cite{10.1063/1.5097141, 10.1063/1.4929790, doi:10.1021/acs.jpca.2c04349} proposes to map the imaginary time path to a two-replica coarse-grained system. 
This mapping gives access to both diagonal and nondiagonal elements of the thermal density matrix. 
However, the corresponding formalism is challenging to implement for realistic molecular systems in the regime of strong NQEs~\cite{doi:10.1021/acs.jpca.2c04349}. 
Another common mapping involves coarse-graining of the ring polymer to the centroid \cite{doi:10.1021/acs.jctc.2c00706, doi:10.1021/acs.jctc.3c00921, Musil2022}.
Amongst others, we have recently shown that~\cite{Musil2022}, the resulting effective potential can be used in the same manner as a classical force field, providing an approximation to the dynamical properties of the system within the centroid molecular dynamics approach~\cite{cao_formulation_1994}. 
However, centroid-based ring-polymer coarse-graining does not give rigorous access to exact static properties~\cite{Cao1994EquilibriumProperties}. 
For instance, for a harmonic oscillator (which is a good approximation for many chemical systems with stiff bonds), the centroid approximation is equivalent to a classical description. 
On the other hand, a prominent feature of our work is the path-integral coarse-grained simulations (PIGS) method that uses many-body universal approximators~\cite{Noid2008} to fit the coarse-grained potential of mean force. 
The PIGS approach utilizes machine-learned potentials, capable of approximating arbitrary complex functions, to learn the full many-body coarse-grained potentials as demonstrated recently on biomolecular simulations~\cite{cgnet,Husic2020,Majewski2023,Charron2023}. \\

Here, instead of dynamical properties, we focus on characterizing static position-dependent properties, including NQEs, and we develop an effective machine-learned potential to compute them accurately and at the cost of classical MD. 
Our approach builds upon the PIGS approach that uses multiscale coarse-graining~\cite{Noid2008} and our work on the use of deep learning architectures for classical coarse-grained force-fields~\cite{KraemerDurumeric2023} but maps the ring polymer onto a single replica instead of the centroid.
This coarse-grained mapping is distinct from earlier choices and allows quantum thermodynamic expectation values of position-dependent observables to be estimated rigorously as simple time averages akin to classical MD. 
We demonstrate that this approach yields accurate quantum nuclear statistics of bulk and isolated systems. 
The remaining sections of this paper are organized as follows: in \cref{seq:Theory}, we briefly summarize the imaginary-time PI theory and the application of thermodynamic coarse-graining to the problem. 
\cref{seq:comp_details} discusses the numeric implementation of the approach for four model systems: one-dimensional Morse potential, single H$_2$O molecule, Zundel cation, and bulk water. The results are presented and discussed in  \cref{seq:results}. \cref{seq:conclusion} provides concluding remarks. 

\section{Theory and Methods}
\label{seq:Theory}
\subsection{The path-integral approach}

We consider a Hamiltonian $H$ of $N$ distinguishable nuclei, living in three-dimensional Cartesian space, with masses $\{m_{1}, \dots, m_{N}\} \equiv \{m_{i}\}_{i}$, and position vector $\mathbf{q} \equiv \{\mathbf{q}_{i}\}_{i}$, interacting with the potential $U(\mathbf{q})$. The equivalent PI partition function comprising $P$ replicas can be expressed as,
\begin{equation}
Z =  \text{tr}\left[e^{-\beta \hat{H}}\right] = \lim _{P\to\infty}Z_P,
\end{equation}
where 
\begin{align}
    Z_P &\propto \int \dint{} \mathbf{Q} ~e^{-\beta U^{\text{PI}}(\mathbf{Q})}, \nonumber \\
    &= \int \prod_{j=1}^{{}} \dint{}\mathbf{q}^{(j)}~ e^{-\frac{\beta}{P}\left[\sum_{j=1}^P \left[\sum_{i}\frac{1}{2} \frac{m_{i} P^2}{\beta^2 \hbar^2} \left(\mathbf{q}_{i}^{(j+1)} - \mathbf{q}_{i}^{(j)}
    \right)^2 + U\left(\mathbf{q}^{(j)}\right)\right]\right]}.
    \label{eq:isomorphism}
\end{align}
Here, $\mathbf{q}^{(j)}$ represents the position vector of the $j$-th replica of the system in the ring polymer with cyclic boundary conditions $\mathbf{q}^{(j)} \equiv \mathbf{q}^{(j+P)}$. In addition, $\mathbf{Q} \equiv \{\mathbf{q}^{(j)}\}_{j}$ is a shorthand for the set of positions of the $P$ replicas of the ring polymer, and $U^{\text{PI}}(\mathbf{Q})$ is a shorthand notation for the total potential experienced by the ring polymer. 
In the limit $P \rightarrow \infty$, the ring polymer statistics is an unbiased estimator of quantum statistics of the target system. \\

The quantum thermodynamic average of a position-dependent operator $O(\hat{\mathbf{q}})$ is typically estimated as the ensemble average of the position-dependent function averaged over all the replicas, 
\begin{align}
    \mathcal{O} = \left\langle \frac{1}{P} \sum_{j=1}^{P}  O(\mathbf{q}^{(j)}) \right\rangle_{\text{PI}},
    \label{eq:ensemble_average}
\end{align}
where $\langle\square\rangle_{\text{PI}}$ corresponds to the ensemble average defined in Eq.~\ref{eq:isomorphism}. 
Estimating Eq.~\ref{eq:ensemble_average} is computationally demanding due to the need for simulating $P$ replicas of the system.
Typically, $P$ should be larger than $\beta\hbar\omega_{\text{max}}$ with $\omega_{\text{max}}$ being the largest physical frequency of the system~\cite{markland_refined_2008}.
Notably, for a given system (with fixed $\omega_{\text{max}}$), the number of replicas scales inversely with $T$, implying a sleepy rising computational cost as the temperature goes to zero.

\subsection{Path-integral coarse-graining for quantum statistics}

Our goal is to reduce the computational cost of simulating ring polymer statistics by integrating additional degrees of freedom, such as (linear combinations of) the replicas of the system, in a way that retains the accuracy of Eq.~\ref{eq:isomorphism}.
In the most extreme and computationally beneficial scenario, we can integrate the $P-1$ replicas of the physical system to an effective potential felt by a single replica, reducing the dimensionality of the ring polymer to the physical system.
In doing so, we reduce the cost of estimating thermodynamic properties to that of classical MD. \\

The problem can be reformulated as follows, using the bottom-up coarse-graining technique widely used in biomolecular simulations.  
We wish to coarse-grain the ring polymer system in the high dimensional configurational space $\mathbf{Q} \in \mathds{R}^{3N P}$ described by a potential $U^{\text{PI}}(\mathbf{Q}$) into an effective thermodynamically consistent potential $U^{\text{eff}}(\mathbf{q})$ with $\mathbf{q} \in \mathds{R}^{3N}$, 
\begin{equation}
\label{eq:pmf}
U^{\text{eff}}(\mathbf{q}) = -\beta^{-1} \ln \left [
\frac{\int \dint{} \mathbf{Q} ~e^{-\beta U^{\text{PI}}(\mathbf{Q})} ~\delta(\mathbf{q}-\xi (\mathbf{Q}))}
{\int \dint{} \mathbf{Q} ~e^{-\beta U^{\text{PI}}(\mathbf{Q})}}
\right],
\end{equation}
where $\delta$ is a Dirac delta function, and $\xi$ is a linear operator that maps coordinates from $\mathds{R}^{3N P} \to \mathds{R}^{3N}$. 
The effective potential of mean force in Eq.~\ref{eq:pmf} can be obtained using the force-matching variational approach~\cite{Noid2008}. 
We first represent the effective potential as a function, $\tilde{U}^{\text{eff}}(\mathbf{q}, \theta)$, parametrized by $\bm{\theta}$, and then minimize the force-matching loss~\cite{cgnet} with respect to $\bm{\theta}$:
\begin{equation}
\mathcal{L} = || \xi_{f}(\mathbf{F}) + \nabla_{\mathbf{q}} \tilde{U}^{\text{eff}} (\mathbf{q}; \bm{\theta}) ||.
\label{eq:forcematching}
\end{equation}
Here, $\mathbf{F} \equiv -\nabla_{\mathbf{Q}} U^{\text{PI}}(\mathbf{Q})$ is the force associated with the (high-dimensional) ring polymer systems, $\xi_f$ is a linear operator that maps the forces from $\mathds{R}^{3N P} \to \mathds{R}^{3N}$, i.e., the high-dimensional ring polymer space into the low-dimensional space associated with the physical system.

\subsection{Single replica coarse-graining}

While several choices of the mapping operators $\xi$ and $\xi_f$ are possible~\cite{KraemerDurumeric2023}, not all choices are ideal for estimating generic position-dependent operators with the full accuracy of Eq.~\ref{eq:ensemble_average}. 
To make a physically meaningful and computationally favorable choice, we note that all replicas of the ring polymer are equivalent thanks to the invariance of the trace in Eq.~\ref{eq:isomorphism} to cyclic shifts.
Hence, a thermodynamic expectation value can be estimated by averaging the position-dependent function estimated on \textit{any} replica $j'$
\begin{align}
    \mathcal{O} = \left\langle O(\mathbf{q}^{(j')}) \right\rangle_{\beta}. 
    \label{eq:ensemble_average_alternate}
\end{align}

Eq.~\ref{eq:ensemble_average_alternate} suggests that the statistics of a single replica are sufficient to estimate equilibrium averages of position-dependent observables. 
From the perspective of developing an effective potential, it is sufficient to integrate all but one (arbitrary) replica. 
Hence, in this work, we coarse-grain the ring polymer to a single replica using as a configurational map the following transformation:
\begin{equation}
\label{eq:coor_transform}
\xi(\mathbf{Q}) = \mathbf{q}^{(j')}.
\end{equation}
Given this configurational map, a thermodynamically consistent force map $\xi_f$ can be constructed with the matrix elements given by following relationship \cite{KraemerDurumeric2023}: 

\begin{equation}
\label{eq:force_map}
\xi_f^{ji} =
\begin{cases}
    1, & \text{for atom } i \text{ identified with replica } j \\
    0, & \text{for atom } i \text{ identified with replica }\; k\ne j \\
    c_{ji} \in \mathbb{R}, & \text{for all other atoms},
\end{cases}
\end{equation}
where $c_{ji}$ are arbitrary.
For a ring polymer, setting $c_{ji} = 0$ leads to the following primitive estimator  of force projected onto coarse-grained space,
\begin{equation}
\label{eq:cg_force}
 \xi_{f}(\mathbf{F}) 
= -\frac{\nabla U(\mathbf{q}^{(j')})}{P} - \frac{mP}  
{\beta^2\hbar^2}(2\mathbf{q}^{(j')} - \mathbf{q}^{(j' + 1)} - \mathbf{q}^{(j' - 1)}).
\end{equation}
However, the numerical efficiency of the force-matching minimization can be further improved by optimizing $c_{ji}$ to minimize the variance of the mapped force \cite{KraemerDurumeric2023}.
With the configurational mapping operator given by \cref{eq:coor_transform}, expectation values of a position-dependent operator can be estimated as a simple ensemble average
\begin{equation}
    \mathcal{O} = \left\langle O(\mathbf{q}) \right\rangle_{\beta}, 
\label{eq:ensemble_average_pigs}
\end{equation}
akin to standard MD, albeit in the classical ensemble of the effective potential ${U}^{\text{eff}}(\mathbf{q})$.  \\

The choice of the mapping operator in Eq.~\ref{eq:coor_transform} is distinct from earlier choices made in the context of coarse-graining of imaginary-time PIs. 
The most popular coarse-graining approach maps the ring polymer positions to its centroid.
While this approach is useful for simulating dynamical properties, it doesn't give correct equilibrium averages of generic position-dependent operators when estimated as an ensemble average similar to Eq.~\ref{eq:ensemble_average_pigs}.
Similarly, the PICG theory introduced in Ref.~\citenum{10.1063/1.5097141, 10.1063/1.4929790, doi:10.1021/acs.jpca.2c04349} coarse-grains the ring polymer to two replicas, which are related to the positions at which the off-diagonal elements of the Boltzmann operator are estimated. 
This approach gives access to generic position- and momentum-dependent operators; however, it requires derivations of bespoke (complicated) operators. 
On the other hand, our mapping reduces the ring polymer to the position at which the trace of the Boltzmann operator is estimated, with ${U}^{\text{eff}}(\mathbf{q}) = - \beta^{-1} \log{ \left<\mathbf{q}\right| e^{-\beta \hat{H}}\left|\mathbf{q}\right>}$,  modulo an additive constant, giving access to position-dependent operators as simple ensemble averages. \\

\subsection{Workflow using the PIGS method}
 
\begin{figure*}[ht]
\includegraphics[scale=1.0]{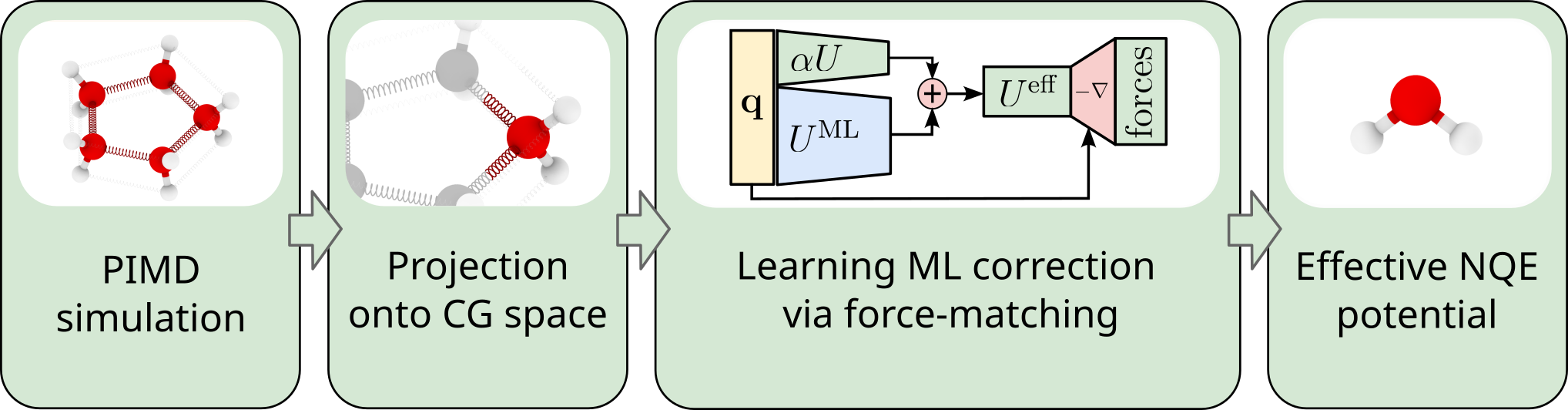}
\caption{Procedure to train an effective NQE potential}
\label{fig:workflow}
\end{figure*}

To develop the effective potentials that incorporate NQEs, we utilize the PIGS approach, outlined in \cref{fig:workflow}.
First, we generate training data via PIMD simulation. The resulting ring polymer trajectories are then projected onto a coarse-grained space using a configurational and force map discussed above. Notice that the remaining replica $j'$ in \cref{eq:coor_transform} is selected arbitrarily, and for a ring polymer with P replicas \cref{eq:coor_transform} defines P equivalent projections. Thus, each ring polymer configuration yields P data points for training. \\

The effective potential is represented in a form 
\begin{equation}
\label{eq:ML_potential}
U^{\text{eff}}(\mathbf{q}, \theta) = \alpha U(\mathbf{q}) + U^{\text{ML}}(\mathbf{q}, \bm{\theta}),
\end{equation}
where $U^{\text{ML}}(\mathbf{q}, \bm{\theta})$ is a learnable correction.
The hyperparameter $\alpha$ determines the contribution of the physical potential to the total energy. 
A  common approach is to set $\alpha = 1$ so the ML network learns an effective NQEs correction to the classical potential \cite{doi:10.1021/acs.jctc.3c00921, Musil2022}. This choice is effective at high temperatures when NQEs are small. However, as temperature decreases, prior classical distribution quickly becomes more localized, and the difference between the classical and target quantum potential increases, making the correction harder to learn. 
However, in the harmonic limit, it is possible to make classical distribution more "quantum-like" by changing the simulation temperature. For this reason, to simplify the learning, enhance the data efficiency of the training, and the model stability, we consider $\alpha$ to be temperature dependent:  $\alpha=T/T_0$, where $T < T_0 $ denotes the target temperature and $T_0$ is a hyperparameter. If not specified otherwise, we set $T_0=600$ K.
The learnable model parameters are then optimized with standard ML approaches. The resulting effective potential can then be used for calculations as a standard MD force field.

\subsection{Applicability of the models to different thermodynamic conditions}
\label{sec:extention}

The thermodynamically consistent CG model approximates the potential of mean force, which depends on the thermodynamic state, e.g., system temperature. 
Consequently, the model is strictly applicable only at the same conditions it was trained, and targeting the coarse-grained model to different conditions requires a different set of training data. 
However, in the case of ring polymer coarse-graining, we notice two cases where the applicability of a trained model can be extended, or training data generated for one condition can be reused for a different one. 
In the first scenario, the molecule predominantly populates the ground vibrational state, and the population of the excited states is negligible. In this case, the general equation for the quantum PMF (see \cref{eq:PMF_exact}) simplifies to:
\begin{equation}
\label{eq:pmf_ground}
U^{\text{eff}} = - \frac{1}{\beta} \ln |\Psi_0|^2,
\end{equation}
where $\Psi_0$ denotes the ground-state nuclear wave function.

Consequently, the effective CG potentials at different temperatures in the ground vibrational state are related to each other as: 
\begin{equation}
\label{eq:temperature_rescaling}
U^{\text{eff}}(\beta_2) = \frac{\beta_1}{\beta_2} U^{\text{eff}}(\beta_1).
\end{equation}
This relationship implies that if a model is trained at a temperature that is low enough to predominately populate the ground state, the same model can be used
to run simulations and compute the quantum statistics at any lower temperature by rescaling $U^{\text{eff}}$ and the corresponding forces according to \cref{eq:temperature_rescaling}. \\

This simple relation breaks down when the population of the excited vibrational states is non-negligible. However, a dataset generated with expensive PIMD simulation at one temperature can be used to generate synthetic datasets corresponding to different temperatures and isotope compositions. 
This data recycling process involves two steps.
First, we reweight the configuration to the target temperature and isotope composition, efficiently using the coordinate rescaling approaches from Refs.~\cite{Yamamoto2005, Ceriotti2013_isotope}. 
Within the coordinate scaling approach, given the initial coordinates $\mathbf{q}$, inverse temperature $\beta_1$ and mass $m_1$, to model a system at inverse temperature $\beta_2$ and mass $m_2$, we rescale the coordinates as
\begin{equation}
\label{eq:rescaling}
 \mathbf{q'}^{(j)} = \bar{\mathbf{q}} - \sqrt{\frac{m_2\beta_1}{m_1\beta_2}}(\bar{\mathbf{q}} - \mathbf{q}^{(j)}),
\end{equation}
where $\bar{\mathbf{q}} = \frac{1}{P} \sum_{j=1}^P \mathbf{q}^{(j)}$ represents the centroid position.
Then, the probability $\rho(\mathbf{q})$ of a configuration in the ensemble characterized by $m_2$ and $\beta_2$ is given by (see SI for the details):
\begin{equation}
\label{eq:weights}
\rho(\mathbf{q'}; \beta_2; m_2) = 
\rho(\mathbf{q}; \beta_1; m_1) \exp \left ( 
\sum_{j=1}^P (\beta_1 \mathbf{q}^{(j)} -  \beta_2 \mathbf{q'}^{(j)}).
\right)
\end{equation}
We used \cref{eq:weights} to reweight the original dataset. 

Second, we recompute the forces originating from the harmonic couplings between adjacent replicas, $ \mathbf{F}'$, for the target temperature and isotope compositions as:
\begin{equation}
 \mathbf{F}'(\mathbf{q}; m_2; \beta_2)  =
  \frac{\beta^2_1 m_2 }{\beta^2_2 m_1}  \mathbf{F}'(\mathbf{q}; m_1; \beta_1).
\end{equation}
The resulting dataset is then used for model training, as described above.\\

\section{Computational details}
\label{seq:comp_details}

\subsection{1-D Morse potential}

\subsubsection{Model and training procedure}

We first demonstrate the approach on a systems for which the exact effective potential can be estimated numerically. 
We study a particle with mass $\mu$ experiencing a one-dimensional Morse potential,
\begin{equation}
\label{eq:Morse}
  U^{\text{Morse}}(q) = D_e(1-e^{-a(q-q_0)})^2,
\end{equation}
where $q$ is the particle position.  The parameters, $q_0$, $a$, and $D_e$, that determine the equilibrium position, shape, and depth of the potential, are chosen to reproduce the potential of the O--H bond, and $\mu$ is set to the reduced mass of the O--H bond (see SI for the numeric values). \\

The effective potential corresponding to the mapping in Eq.~\ref{eq:coor_transform} can be estimated numerically by solving the  Schr\"{o}dinger equation for the bound states of a Morse potential~\cite{10.1063/1.453761} as
\begin{equation}
   \label{eq:PMF_exact}
    U^{\text{eff}}(q) = -\beta^{-1}\ln{}\frac{\sum_i e^{-\beta E_i}|\Psi_i(q)|^2}{\sum_i e^{-\beta E_i}},
\end{equation}
where $E_i$ and $\Psi_i(q)$ are the 
the energy eigenvalues and wave functions associated with the $i$-th eigenstate, respectively. 
To estimate the effective potential using the PIGS approach, we optimize a coarse-grained potential in the following \cref{eq:ML_potential} with $\alpha=P^{-1}$: 
\begin{equation}
\label{eq:Morse_CG}
U^{\text{eff}}(q, \bm{\theta})  = \frac{U^{\text{Morse}}(q)}{P} + U^{\text{ML}}(q, \bm{\theta}),
\end{equation}
where $U^{\text{ML}}= \sum_{i=1}^M \theta_i ~\phi_i(q)$ is a linear combination of radial basis functions $\phi_i(q)$ (see SI for details)~\cite{doi:10.1021/acs.jctc.9b00181}, and $M$ is the total number of basis functions (we set $M=10$).
Hence, instead of learning the full effective potential, we fit a correction to the classical potential that captures the quantum effects learned from the full-ring polymer simulations.
The adjustable parameters $\bm{\theta}$ are obtained by minimizing \cref{eq:forcematching} using ridge regression, with the $L_2$  regularization parameter set to 0.1 and 5-fold cross-validation.

\subsubsection{Training data generation}
The training data was generated from PIMD simulations using the \texttt{i-PI} code~\cite{KAPIL2019214} and a custom Python-based force provider implementing the one-dimensional Morse potential. 
Six simulations were performed in an NVT ensemble for 100 - 600\,K in steps of 100\,K, using 64 replicas and a PILE-L thermostat with a time constant of $100$\,fs~\cite{10.1063/1.3489925} 
The integration was performed with a timestep of $0.5$ fs and positions and forces acting on each bead were sampled every $20$ steps. 
The first $2.5$ ps of the simulation were discarded, and the positions and forces sampled in the subsequent $500$ ps were used to train the model.

\subsection {Molecular systems: water molecule, Zundel cation and liquid water}

\subsubsection{Model and training procedure}
We use \cref{eq:ML_potential} to model the effective potential of the single water molecule, the Zundel cation, and bulk water.
We represent $U^{\text{ML}}$ with the MACE graph neural network architecture~\cite{batatia2023mace}. 
One of the hallmarks of the MACE architecture is the use of many-body features to represent atomic local environments, improving the model accuracy and data efficiency. \\
%

The models were trained with the AdamW optimizer, a learning rate of $0.001$, and a weight decay of $0.001$. 
The MACE neural network cutoff was set to $6.0$ \AA, with $15$ radial basis functions,  $6$ polynomial, and $3$ radial channels. 
The correlation order for Zundel cation and bulk water was set to $3$, which corresponds to $4$-body features. 
For a single water molecule,  $3$-body features were used. 
In all cases, 
$80$ \% of the data available were used for training, and the remaining $20$ \% were used for validation.
The training was performed for $60$ epochs for the H$_2$O molecule and the Zundel cation and for $15$ epochs for liquid H$_2$O.
For each system, ten different models were trained, each using different parameter initialization and train-validation splitting. 

\subsubsection{Training data generation}
The training data was generated from PIMD simulations using the \texttt{i-PI} code using the Partridge and Schwenke potential for a single water molecule~\cite{10.1063/1.473987},  the HBB potential~\cite{10.1063/1.1834500} for the Zundel cation and the MB-pol for bulk water  \cite{doi:10.1021/ct400863t}. The simulation parameters are summarized in Table S1. 
We use a force map optimized to minimize the average magnitude of the mapped forces, subject to the constraints of the valid force map. This force optimization was performed according to the procedure introduced in ~\cite{KraemerDurumeric2023}, with an $L_2$ regularization parameter set to $0.05$. 

\subsubsection{Model evaluation}
For each of the trained models, we performed molecular dynamics simulations, using as a force field a combination of classical potential and a trained correction as defined in Eq.~\ref{eq:ML_potential}. The simulations were performed in the NVT ensemble, with a Langevin integrator and a friction thermostat with a time constant  set to $100$ ps$^{-1}$. The time step and periodic boundary conditions were the same as in the training simulations. All the frames were used for further analysis. 
For each of the systems, we performed additional simulations with only the classical potential as a control to compare the classical and quantum statistics. \\
\section{Results and Discussion}
\label{seq:results}

\subsection{1-D Morse potential}

\begin{figure*}[t!]
\includegraphics[scale=1.0]{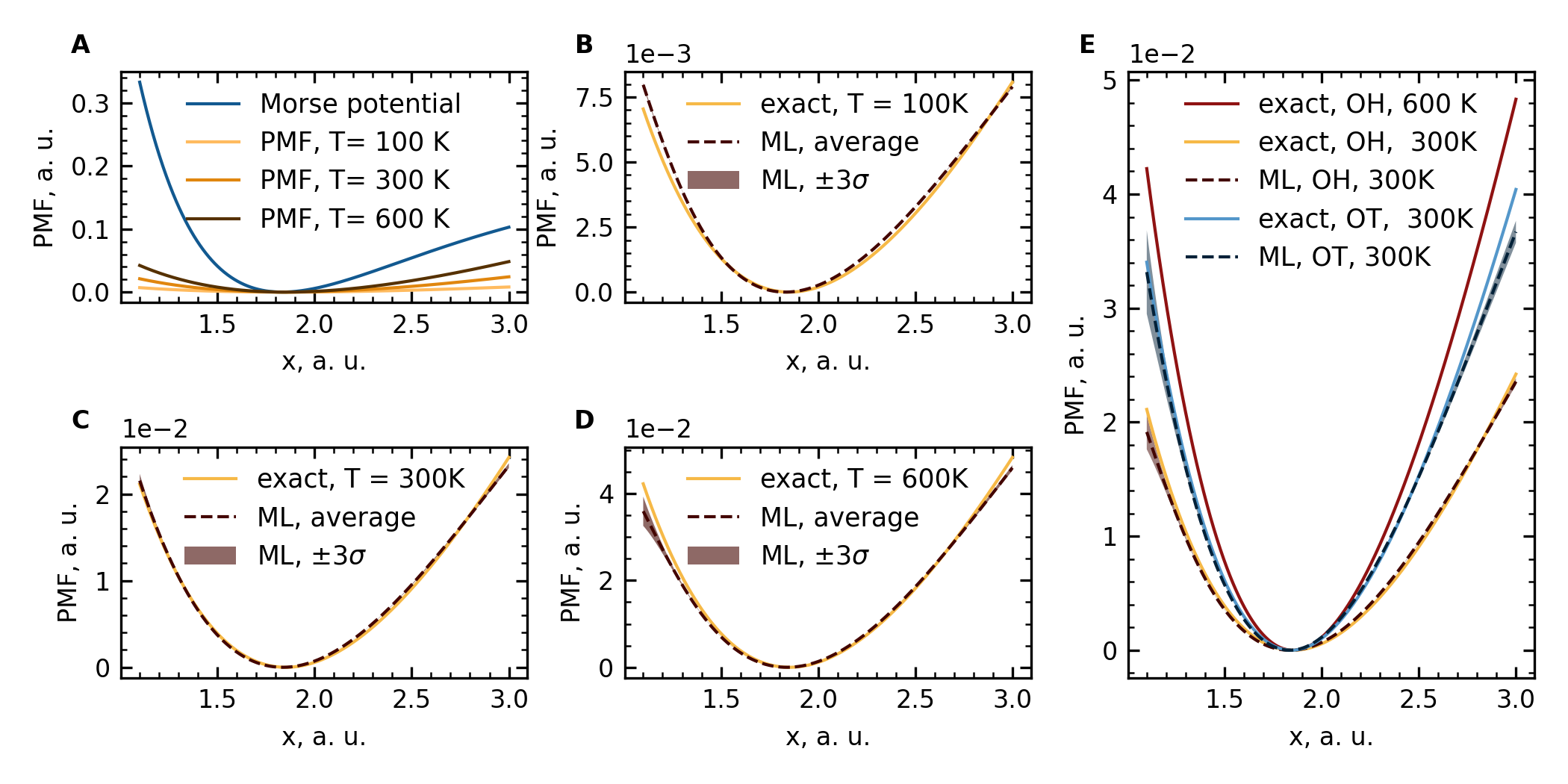}
\caption{Exact and machine-learned PMF for the Morse potential. Panel A: Classical Morse potential and quantum PMFs at different temperatures. Even at $600$K, the system exhibits significant quantum effects.
Panels B-D: comparison of exact quantum PMF and machine-learned PMF at $100$K (B), $300$K (C), and $600$K (D).   
Panel E: Results for models trained at 300 K for OH and OT from reweighted simulations performed at $600$K.
The average PMF over 5 models is reported as a dashed line, and  $\pm$ 3 standard deviations are reported as shaded areas }
\label{fig:Morse}
\end{figure*}

First, we study a particle in a one-dimensional Morse potential, a model for a single O--H bond.  
The Schr\"{o}dinger equation for this system can be solved analytically \cite{10.1063/1.453761}, thus making Morse potential an ideal test system. 
It also exhibits significant nuclear quantum effects, as shown in Fig. \ref{fig:Morse}A where the classical Morse potential (\cref{eq:Morse}) is compared with the corresponding quantum PMFs (\cref{eq:PMF_exact}) at different temperatures. The classical approximation leads to the over-localization of the particle, and, as expected, the quantum effects are more prominent at low temperatures. However, for this system, even at a temperature of $600$K, there are significant deviations between the classical and quantum pictures. 

The developed approach allows us to get a quantitative agreement with the reference, especially in the high-probability regions that are well-sampled in the training dataset (Fig. \ref{fig:Morse}B-D).  Further away from the minimum energy, the model's prediction slightly deviates from the ground truth. This is to be expected, as these regions of the phase space are not as well represented in the training data as the high-probability regions.  

\begin{figure*}[ht]
\includegraphics[scale=1.0]{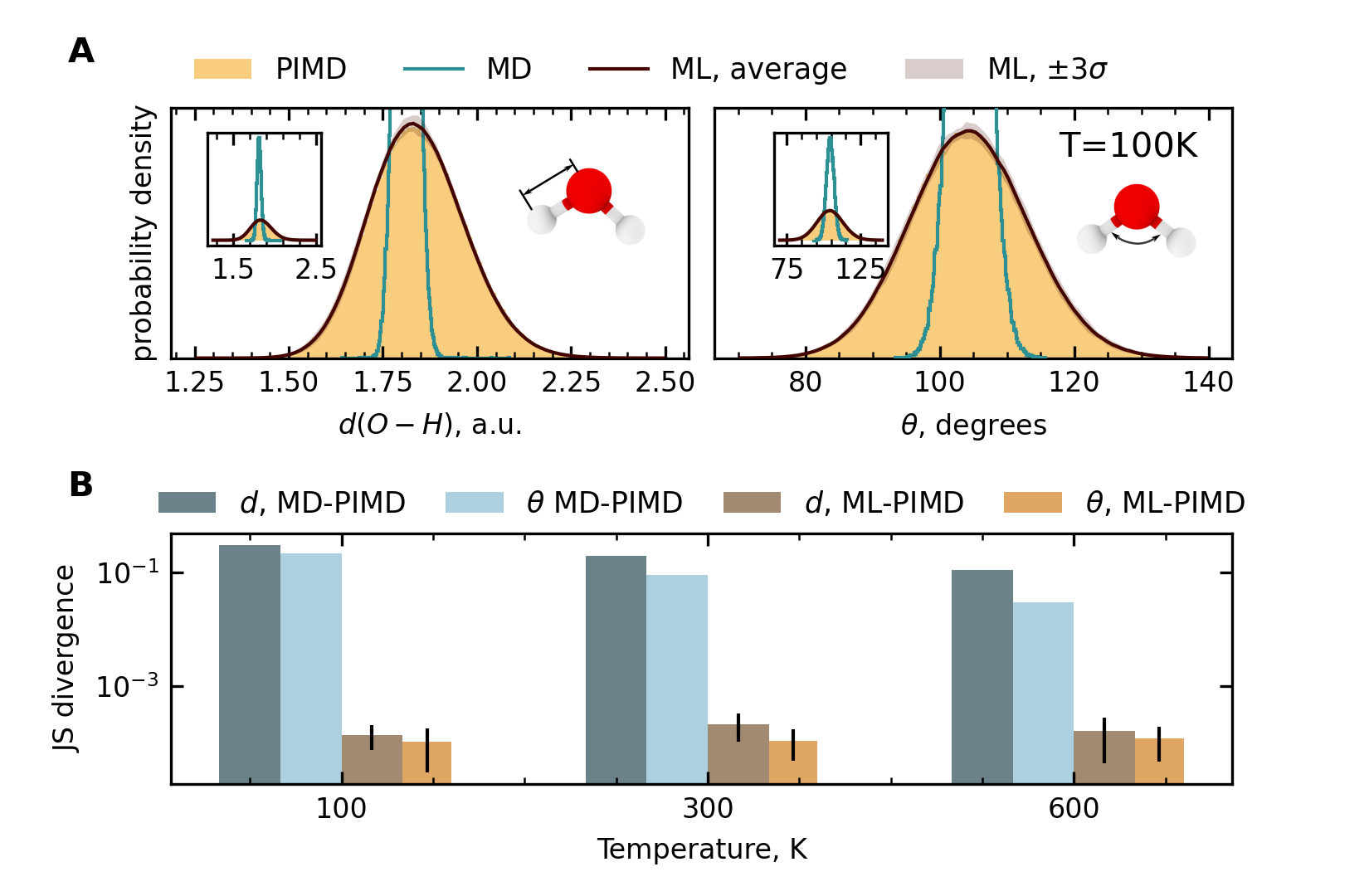}
\caption{A: Comparison of the probability distributions of the O$-$H bond length (left) and the H$-$O$-$H angle (right) for classical MD (P=1), PIMD (P=64), and ML models at 100 K. Inserts show the same distributions at a different scale. 
B: Jensen-Shannon (JS) divergence between collective variable distributions obtained with PIMD and either classical potential (shades of blue) or with our machine-learned model (shades of brown). For each case, JS divergence for the O$-$H bond length distribution and the H$-$O$-$H angle at three different temperatures are shown. }
\label{fig:H2O_mol}
\end{figure*}

We test the reweighting scheme proposed in \cref{sec:extention} by generating datasets corresponding to different temperatures ($300$ K) and mass of the particle that is either Protium   (corresponding to an O-H bond) or increased to Tritium  (corresponding to an O-T bond) from the $600$K O-H dataset previously used.
The resulting reweighted datasets were used to train new models. 
As shown in Fig. \ref{fig:Morse}E, the models trained with the reweighted dataset are in excellent agreement with the analytical results for the corresponding temperature and isotope composition. Such data recycling allows for a further decrease in the computational cost of generating training data, though it becomes less effective for more complex systems.
Nevertheless, we expect that this approach may be particularly useful for generating large datasets for training transferable models. 

\subsection {Water molecule}

The next test case we considered is a single water molecule in a vacuum.  The quantum statistics of this 3-body system can be exhaustively described in terms of the length of two O$-$H bonds and the H$-$O$-$H bond angle. The comparison of the probability distributions for the O$-$H distances and H$-$O$-$H bond angles is shown in \cref{fig:H2O_mol}, with additional details provided in Fig. S1. As was the case for the Morse potential, the classical distribution is significantly more localized than the quantum counterpart. The ML model consistently samples the correct quantum distribution: the results obtained from 10 independently trained models produce similar results with low variance. \\

\begin{figure*}[ht]
\includegraphics[scale=1.0]{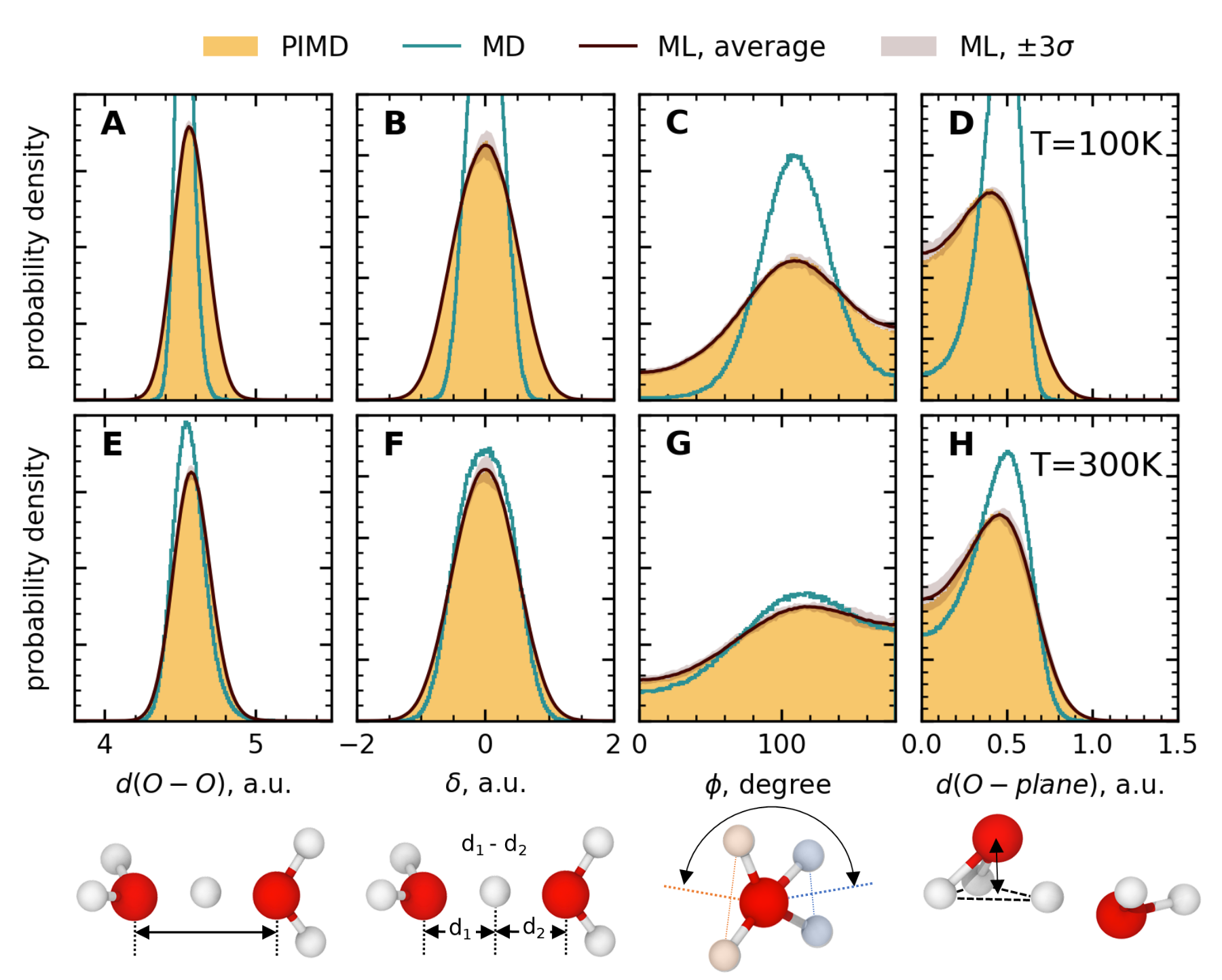}
\caption{The probability distribution for the Zundel cation, sampled with PIMD simulation, MD simulation, and the proposed ML model. The top row (panels A-D) represents the model trained and simulated at $100$K, and the bottom row (panels E-H) shows the results trained at $300$K. Under the plots, a visual representation of the corresponding collective variable is shown.}
\label{fig:Zundel}
\end{figure*}

For all trained models, we run the MD simulation for one ns, which is four times longer than the PIMD simulation used to generate the training data. This allows us to obtain slowly converging properties by performing short PIMD simulations and using those to train a model that can be run for a much longer time. 

Even at $600$K, the water molecule predominantly populates the ground vibrational state, with a negligible population of the excited states. Thus, as discussed in \cref{sec:extention}, the model trained at $600$K can be used to simulate the system at any lower temperatures.  To demonstrate this point further, we report in Fig. S2 the results obtained with a model trained on PIMD data generated at $600$K and then simulated at lower temperatures, with forces rescaled according to \cref{eq:temperature_rescaling}.

\subsection{Zundel cation}

A more challenging case is the Zundel cation, which consists of two water molecules sharing a hydrogen-bonded proton H$^*$. This system has been extensively studied both theoretically \cite{Schroeder2022,Suzuki2013} and experimentally \cite{Headrick2004, Diken2005, Hammer2005, McCunn2008, Vendrell_Gatti_Meyer_2008}. The Zundel cation exhibits large conformational fluctuations, including anharmonic and floppy motion associated with the transfer of the hydrogen-bounded proton.  
Following Ref.~\cite{Suzuki2013}, we characterize these fluctuations by means of the O$-$O distance  $d(\text{O}-\text{O})$, the proton transfer coordinate $ \delta = d(\text{O}_1\text{H}^*) -d(\text{O}_2\text{H}^*) $,   the dihedral angle $\phi$ that measures the rotation of the H$_3$O$^+$ fragments around the O$-$H$^*-$O axis, and the $d(\text{O}-\text{plane})$ that measures the nonplanarity of the H$_3$O$^+$ fragment. The dihedral angle $\phi$  is defined by points X$_1 - $O$_1  - $O$_2 -$X$_2$, where X$_1$ and X$_2$ represent a midpoint between the hydrogen atoms that are not involved in the hydrogen bond formation and bound to O$_1$ and O$_2$, correspondingly. The non-planarity of H$_3$O$^+$ is characterized by a distance between an O atom and the plane, defined by the three closest H atoms (see the lower panel of \cref{fig:Zundel}). 

\begin{figure*}[ht]
\includegraphics[scale=1.0]{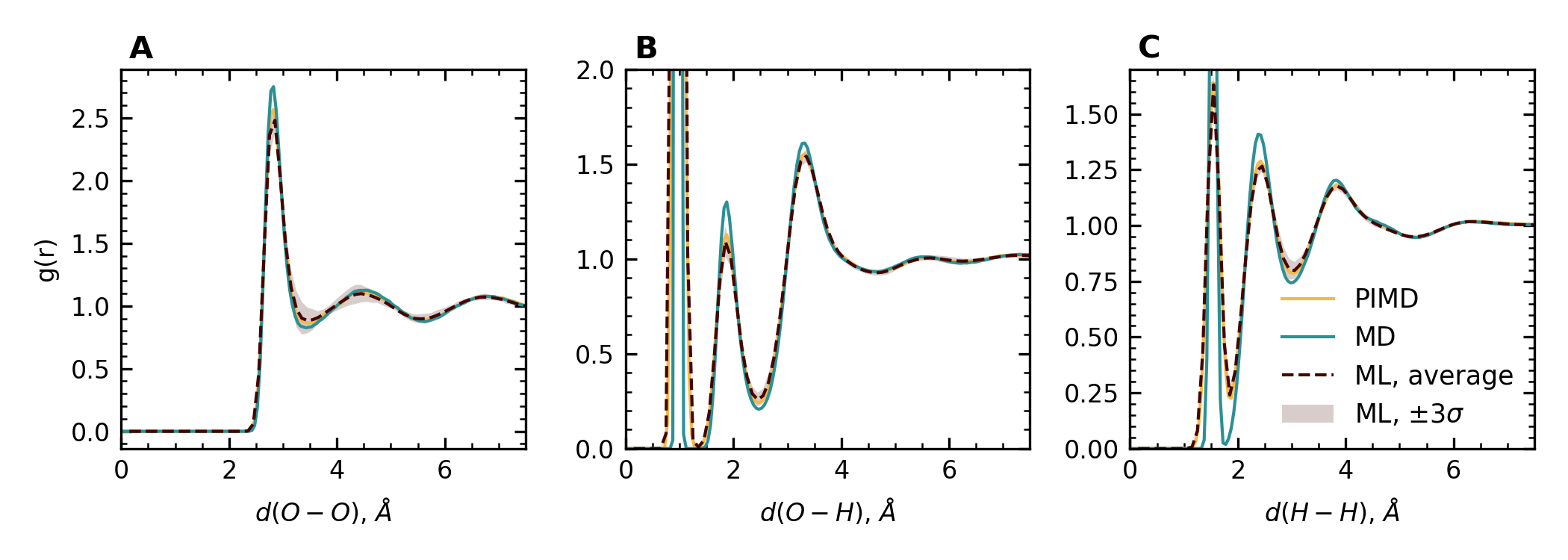}
\caption{Radial distribution functions $g(r)^{OO}$, $g(r)^{OH}$, and $g(r)^{HH}$ for bulk H$_2$O at $300$ K. 
The teal line represents the RDF obtained with a classical MD simulation (P=1), the orange line represents the PIMD result, and the dashed brown line represents the results obtained with the ML model. The result reported for the ML model is the mean over ten independent models with different train-test splits and model initialization.}
\label{fig:bulk_h2o}
\end{figure*}

At $300$K, the proton transfer coordinate and the rotational motion are almost classical, and only a small correction is needed to account for NQEs. The O$-$O distance and the O$-$plane distance require larger corrections, yet the NQEs are relatively small. At $100$K, however, significant quantum effects appear in all the collective variables, including changes in the equilibrium geometry. Nevertheless, our methods recover the accurate quantum distribution both at $300$K and at $100$K.
Despite the diversity of conformational fluctuations, the model not only reproduces the target probability distributions but leads to remarkably stable simulations: all of the models were used to perform $1$ ns simulation and did not exhibit any instabilities, often plaguing ML force fields~\cite{fu2023, Stocker2022}.  \\

\subsection{Liquid water}

Lastly, we applied our approach to model liquid water to reproduce the NQEs manifesting in both bonded and nonbonded interactions.  Since NQEs are highly local, the former are more affected than the latter. The quality of the machine-learned model is evaluated in \cref{fig:bulk_h2o} by computing radial distribution functions (RDFs). 
At $300$K,  the first peak in the O$-$H~(\cref{fig:bulk_h2o}B) and  H$-$H ~(\cref{fig:bulk_h2o}C) RDFs predicted by classical MD (P=1) are too narrow and significantly deviate from the correct quantum PIMD results. The deviations between the classical and quantum RDFs rapidly decrease with the distance.  

Our neural-network-based potential is able to accurately predict the correct quantum RDFs, including the contributions that arise from both bonded and nonbonded interactions~(\cref{fig:bulk_h2o}), in contrast to another similar study \cite{doi:10.1021/acs.jctc.3c00921}. There is also a remarkable agreement between independently trained models. 
The high accuracy and low uncertainty of our ML model prediction show that it reliably captures the subtle differences between classical and quantum RDFs.

Compared to the simulations for the ML models of a single water molecule and of the Zundel cation, the simulations for bulk water exhibit a decreased stability. 
We note that some simulations of our ML models, trained on $35$ ps of PIMD simulations,  exhibit numerical instabilities. However, the ML models allow, on average, at least $38$ ps of simulation time before the onset of any instability. Simulations can be reinitialized and restarted to improve statistics.
The numerical instability of the model can be potentially cured by increasing the size of the training dataset. Alternatively, one can modify the learning problem by simplifying the ML correction, for example, by introducing physics-informed energy terms, as it is done in the learning of classical ML coarse-grained force-fields~\cite{cgnet,Husic2020,Majewski2023,Charron2023}. 
Given that the stability of the simulation strongly depends on the parameter $\alpha$ (see~\cref{eq:ML_potential}), this approach may be especially promising. 
\section{Conclusion}
\label{seq:conclusion}
We developed an approach to coarse-grain the ring-polymer dynamics describing NQEs and we obtain results thermodynamically consistent with PIMD simulations. This approach allows us to accurately compute position-dependent NQEs at the cost of classical MD. We have demonstrated this method on four distinct systems where NQEs are prevalent — the Morse potential, a single water molecule, the Zundel cation, and bulk water - and reported results in excellent agreement with (much more expensive) reference PIMD simulations. 

The proposed method builds an effective NQEs correction potential by combining the rigorous framework of multiscale coarse-graining~\cite{Noid2008} with the MACE architecture~\cite{batatia2023mace}, an expressive state-of-the-art graph neural network potential.
We have shown that these effective potentials accurately reproduce the quantum statistics of nuclei, both near the classical limit and in the regime of strong NQE.
Moreover, our method eliminates the need to simulate multiple replicas.
Once parametrized, the model seamlessly integrates into existing classical force fields and doesn't require specialized algorithms other than ones already used for classical MD simulations.
However, this approach is limited to reproducing quantum thermodynamics, so dynamical and momentum-dependent properties should be approximated using a previously proposed method~\cite{Musil2022}.

For the results presented here, different sets of training data were used to train models for different molecular species at different temperatures. However, if the temperature associated with the training data is low enough to significantly populate only the quantum ground state, the trained model can be used to also simulate any lower temperature, as we have demonstrated in the case of the H$_2$O molecule. The proposed method is designed such that the forces originating from the spring coupling between the adjacent replicas have a non-zero contribution to the force projected onto the coarse-grained space. As the dependence of these forces on the atomic mass and temperature is known (\cref{eq:cg_force}), the training dataset obtained for one isotope composition and temperature can be, in principle, reweighted to construct a dataset for different isotope compositions and temperatures. Thus, a single PIMD simulation can provide data for training several models, as we have shown in the case of the Morse potential.

The rapid proliferation of machine-learned force fields parametrized using accurate quantum chemical calculations to account for the electronic effects means the missing NQEs are now becoming a dominant source of error. 
At the same time, NQEs are crucial to describe biomolecular processes accurately, for instance, in the case of proton transfer reactions. While the results presented here are model-specific, building upon the recent development of machine-learned force-fields (either coarse-grained~\cite{cgnet,Husic2020,Majewski2023,Charron2023}, or atomistic~\cite{Schutt_Sauceda_Kindermans_Tkatchenko_Muller_2018, doi:10.1021/acs.jctc.9b00181, Unke_Chmiela_Gastegger_Schutt_Sauceda_Muller_2021, Unke_2021, Batzner2022,  batatia2023mace}), we believe this study represents a significant step forward towards the design of transferable potentials that explicitly incorporate also NQEs but retain the computational cost of classical MD. 

\section*{Supplementary material}
Section I of the supplementary material details the Morse potential (I.A) and ML model used for a particle in Morse potential (I.B). Section II contains the simulation parameters used to generate the training data.  Section III provides the derivation of coordinate rescaling. Section IV presents extended results for a water molecule.
\section*{Acknowledgements}
We thank members of the Clementi's group at FU for insightful discussions and comments on the manuscript. 
We gratefully acknowledge funding from the Deutsche Forschungsgemeinschaft
DFG (SFB/TRR 186, Project A12; SFB 1114, Projects B03, B08, and A04; SFB 1078, Project C7), the National Science Foundation (PHY-2019745), and the Einstein Foundation Berlin (project 0420815101), and the computing time provided on the supercomputer Lise at NHR@ZIB as part of the NHR infrastructure (project beb00040),  the HPC Service of FUB-IT, Freie Universität Berlin, the HPC Service of the Physics department, Freie Universität Berlin, and the Swiss National Supercomputing Centre (under projects s1209 and s1288). V.K. acknowledges support from the Ernest Oppenheimer Early Career Fellowship and the Sydney Harvey Junior Research Fellowship, Churchill College, University of Cambridge. 
\section*{Author declarations}
\subsection*{Conflict of Interest}
The authors have no conflicts to disclose.

\section*{Data availability}
All the training data, code, and scripts required to reproduce the results are made available in the online repository: {\href{https://github.com/ClementiGroup/accurate_nuclear_quantum_statistics_on_machine_learned_classical_effective_potentials.git}{\texttt{cg\_nuclear\_quantum\_statistics}}}.


%

\end{document}